\documentclass[a4paper,11pt]{article}
\pdfoutput=1 

\usepackage{jcappub} 
\usepackage[T1]{fontenc}

\usepackage{amssymb,amsmath,amsfonts,empheq,amsbsy,epsfig,wrapfig,caption,float,slashed,graphicx, pstricks}
\usepackage{subcaption}
\usepackage{units}
\usepackage{todonotes}

\usepackage{tikz-feynman}
\tikzfeynmanset{compat=1.1.0}
\usepackage{feynmp}
\newcommand{\mpl}{M_\text{Pl}}
\newcommand{\sm}{\text{SM}}
\newcommand{\GW}{\text{GW}}

\usepackage{cancel}
\usepackage[scr=boondox,  
            ]   
           {mathalpha}

\graphicspath{{./plots}}

\title{\boldmath Primordial Gravitational Wave Probes of Non-Standard Thermal Histories}
\author[a]{A. Konings,}
\author[b, 1]{M. Marinichenko,\note{Corresponding author.}}
\author[a]{O. Mikulenko}
\author[a]{and S. P. Patil}

\affiliation[a]{Instituut-Lorentz for Theoretical Physics, Leiden University, \\2333 CA Leiden, The Netherlands}
\affiliation[b]{Leiden Observatory, Leiden University,
\\ 2333 CC Leiden, The Netherlands}
\emailAdd{marinichenko@strw.leidenuniv.nl}
\date{\today}

\abstract{
Primordial gravitational waves propagate almost unimpeded from the moment they are generated to the present epoch. Nevertheless, they are subject to convolution with a non-trivial transfer function. Within the standard thermal history, shifts in the temperature-redshift relation combine with damping effects by free streaming neutrinos to non-trivially process different wavelengths during radiation domination, with subsequently negligible effects at later times. Presuming a nearly scale invariant primordial spectrum, one obtains a characteristic late time spectrum, deviations from which would indicate departures from the standard thermal history. Given the paucity of probes of the early universe physics before nucleosynthesis, it is useful to classify how deviations from the standard thermal history of the early universe can be constrained from observations of the late time stochastic background. The late time spectral density has a plateau at high frequencies that can in principle be significantly enhanced or suppressed relative to the standard thermal history depending on the equation of state of the epoch intervening reheating and the terminal phase of radiation domination, imprinting additional features from bursts of entropy production, and additional damping at intermediate scales via anisotropic stress production. In this paper, we survey phenomenologically motivated scenarios of early matter domination, kination, and late time decaying particles as representative non-standard thermal histories, elaborate on their late time stochastic background, and discuss constraints on different model scenarios.
}

\begin{document}

\maketitle
\flushbottom

\section{Preliminaries}

Gravitational waves offer an almost uninterrupted view of the physics of the very early universe \cite{Maggiore:1999vm, Caprini:2018mtu}. Unlike electromagnetic probes, gravitational waves interact very weakly with matter. This makes them harder to detect but also limits the propensity for intervening matter along the line of sight (i.e. foregrounds) to obscure our view of their source. Although the scattering and absorption of gravitational waves during matter domination (MD) are feeble \cite{Baym:2017xvh, Flauger:2017ged}, free streaming neutrinos have been shown to dissipate power via anisotropic stresses for sub-horizon modes of any stochastic background present during radiation domination (RD) \cite{Weinberg:2003ur}. Furthermore, changes in the temperature-redshift relation due to changing numbers of relativistic species, in addition to any entropy generating processes such as the quantum chromodynamics (QCD) crossover, conspire to process different wavelengths in a characteristic manner via a non-trivial transfer function \cite{watanabe2006improved, Kite:2021yoe}. The result is a late time stochastic background with a power spectral density with various features that encode the particle content and thermal history of the early universe.\footnote{In principle, the transfer function would also have to account for lensing effects from the large scale structure foreground. We neglect this in what follows as we are primarily interested in scales for which this effect is negligible, although a more thorough treatment would certainly have to incorporate these effects \cite{Jenkins_2018, Cusin_2019, Cusin_2017, Contaldi_2017, Bartolo_2020, Bartolo_2022, Valbusa_Dall_Armi_2021, Cusin_2018, Bertacca_2020, Garoffolo:2022usx, Balaudo:2022znx, Braga:2024pik}.}

Any deviations from the thermal history of the universe will propagate through the transfer function, whether this be through modifications to the particle spectrum of the universe, or through the introduction of any additional entropy generating mechanisms (e.g. \cite{Asaka:2020wcr,Berbig:2023yyy,  Borboruah:2024eha, Miron-Granese:2020hyq, Gu:2023fqm, DEramo:2019tit, Dolgov_2011, Dong_2016, Buchmuller:2013dja, Athron:2024fcj, barman2025testingleptogenesisdarkmatter} for a partial survey), as well as possible bursts of anisotropic stress production. Given the limited number of probes that allow us to access the physics of the early universe before Big Bang nucleosynthesis (BBN), primordial gravitational waves appear to offer a constraining lever not accessible to other probes. 

In what follows, we explore the possibility of constraining non-standard thermal histories via observations of the late time stochastic background, taking for granted a nearly scale invariant spectrum of primordial gravitational waves. We do this by surveying a range of phenomenological models where the epoch between reheating and the terminal phase of radiation domination that onsets before nucleosynthesis may not correspond to pure radiation domination over its entirety. There are, of course, plausible scenarios in which the primordial spectrum departs from scale invariance. In what follows, we make the minimal assumption that this is not the case. Although any transfer function formalism is agnostic to the form of the input spectrum (provided it remains perturbatively small), one would not be able to cleanly distinguish spectral deviations induced by modified initial spectra from departures from the standard thermal history without additional observational input.

There are several ways in which one or more phase of non-radiation-like evolution could punctuate the epoch between reheating and nucleosynthesis. The field responsible for reheating (whether inflaton itself or some other intermediary) may undergo non-trivial dynamics around or before pre-reheating before it decays into Standard Model (SM) particles, which then go on to thermalize. There could be additional light fields, such as axions which are frozen up until the Hubble scale drops below their mass, at which point they undergo coherent oscillations with a time averaged equation of state of cold dark matter \cite{sahni2000new, Hu:2000ke, Hui:2016ltb}, which may themselves produce additional light mediators and radiation \cite{Machado:2018nqk, Machado:2019xuc, Madge:2021abk}. One could also envisage departures from an equation of state of $w = 1/3$ from (dark sector) phase transitions, or the decay of heavy long lived particles. Although what is presented in the study that follows represents only a partial survey, it is straightforward to intuit how the dynamics of any given model directly translate into the power spectral density for the stochastic background: In the absence of any other source of anisotropic stress other than neutrinos\footnote{Photons, for example, are also a source of anisotropic stress. However, they are energetically subdominant during radiation domination and can be safely neglected \cite{watanabe2006improved, Kite:2021yoe, Chluba:2014qia}. If the particle content were to include additional light, weakly interacting particles (interactions rapidly dissipate anisotropic stress \cite{Baym:2017xvh}), one would have to factor in their damping effects as well. This can be the case in scenarios involving dark photon production, as we discuss further.}, deviations from an equation of state parameter $w = 1/3$ translate into a suppression or enhancement of the high frequency plateau of the spectral density. Suppression occurs when $\omega$ decreases, and enhancement occurs when it increases in the intervening epoch(s) \cite{Soman:2024zor}, with deviations at earlier times imprinting at shorter comoving scales, or higher frequencies, upon which features are superposed in the usual manner from shifts in the temperature-redshift relation and any intervening bursts of entropy production. If there are mechanisms to generate bursts of anisotropic stress production, then additional damping at intermediate scales also becomes a possibility. 

In the sections that follow, we consider specific models of early matter domination (EMD), long-lived feebly interacting particles, and intervening kination as illustrative examples in which the mechanisms detailed above feature. We subsequently discuss prospects for constraining non-standard thermal histories with future and planned observations before offering our concluding thoughts. We discuss various details about our numerical implementation as we encounter them, referring to previous literature for the various derivations upon which this investigation relies.  

\section{The gravitational wave transfer function}

In what follows, we work in terms of the transfer function $\mathcal{T}(\tau, k)$ that describes the non-trivial evolution of gravitational wave modes $h_{\mathbf{k}}(\tau)$:
\begin{equation}
	h_{\mathbf{k}}(\tau) \equiv h_{\mathbf{k}}^{\text{prim}} \mathcal{T}(\tau, k),
\end{equation}
where $\tau$ denotes the conformal time; $k=|\mathbf{k}|$ is the comoving wavenumber, and $h_{\mathbf{k}}^{\text{prim}}$ is the primordial gravitational wave (PGW) amplitude shortly after horizon crossing during inflation. For a given comoving scale $k$, the equation for transfer function $\mathcal{T}(\tau, k) \equiv \mathcal{T}_k(u)$ can be written in terms of the dimensionless parameter $u = k\tau$ and in presence of anisotropic stress is given by \cite{Weinberg:2003ur, watanabe2006improved}:
\begin{equation}
	\label{eq:final_equation_text}
	\mathcal{T}_k''(u)+2\frac{a'}{a }\mathcal{T}_k'(u)+\mathcal{T}_k(u)=-24f_{\nu}(u)\bigg(\frac{a'}{a}\bigg)^2\int^{u}_{u_{\text{dec}}} \frac{j_2 (u-s)}{(u-s)^2}\mathcal{T}_k'(s) ds,
\end{equation}
with primes denoting the derivative with respect to $u$. The value $u_{\text{dec}}$ corresponds to the moment of neutrino decoupling, and the neutrino fraction $f_{\nu}$ of the total energy density,
\begin{equation}
	\label{eq:neutrino_fraction_evol}
	f_{\nu}(\tau)=\frac{\Omega_\nu}{\Omega_\gamma+\Omega_\nu} \left[1+\frac{a(\tau)}{a_{\text{eq}}}\right]^{-1} ,
\end{equation}
is equal to $0.41$ during radiation domination and gradually decreases after matter-radiation equality, which happens at the scale factor $a_\text{eq}$. Finally, $j_2(u)$ is the spherical Bessel function of the second kind. More details of the derivation that led to the above can be found in \cite{Weinberg:2003ur, watanabe2006improved}.

The complexity of the integro-differential Eq.~\eqref{eq:final_equation_text} requires building the solution independently for each $k$.
We adopt an iterative approach: the resulting function $\mathcal{T}_k(u) = \sum_{n=0}^{N}\mathcal{T}_k^{(n)}$ is computed starting with the solution $\mathcal{T}^{(0)}(u)$ of the homogeneous part of Eq.~\eqref{eq:final_equation_text}, and then finding the corrections $\mathcal{T}^{(n)}(u)$ obtained when the previous term $\mathcal{T}^{(n-1)}(u)$ is substituted into the integral. The initial conditions for every correction are given by:
\begin{equation}
	\mathcal{T}^{(0)}_k(0) = 1, \qquad [\mathcal{T}^{(0)}_k]'(0) = 0,
\qquad	\mathcal{T}_k^{(n)}(0) = 0, \qquad [\mathcal{T}^{(n)}_k]'(0) = 0, \qquad n=1,2,\dots
\end{equation}
Here, $N$ denotes the number of iterations. 

For our purposes, we set the required precision of the method at the percent level, which is achieved with $N \lesssim 10$. 

The differential operator of the l.h.s. of Eq.~\eqref{eq:final_equation_text} leads to oscillatory behavior similar to that of the Bessel functions, making numerical integration up to large $u$ computationally expensive. To deal with this, we integrate up to some sufficiently large cut-off $u_\text{end}=200$ that covers many periods of oscillations, and match the solution $\mathcal T_k(u)$ for larger $u$ with the WKB approximation \cite{watanabe2006improved, Saikawa:2018rcs, Kite:2021yoe}: 
\begin{equation}
	\label{eq:transfer_funct_WKB}
	\mathcal{T}_k(u)=A(u)\sin{(u+\delta)}, \qquad A(u)=A_0\frac{a( \tau = u_\text{end}/k)}{a(\tau = u/k)}.
\end{equation}
The amplitude $A_0$ can be extracted by matching the numerical solution to the analytic formula at the end of the integration range: 
\begin{equation}
	A_0 = \sqrt{[\mathcal{T}(u_\text{end})]^2+[\mathcal{T}'(u_\text{end})]^2}.
\end{equation}
The standard expression for the PGW spectral energy density of $\Omega_{\GW}(\tau,k)$ is then given by~\cite{watanabe2006improved}:
\begin{equation}
	\label{eq:spectrum_analytical}
	\Omega_{\GW}(\tau, k) = \frac{\Delta^2_{\GW, \text{prim}}}{12 H^2(\tau) a^2}  \left[k \mathcal{T}_k'\right]^2,
\end{equation}
where the normalization factor $\Delta^2_{\GW,\text{prim}}$ is defined by the primordial input spectrum. Presuming an epoch of slow-roll single-field inflation to have preceded reheating, the amplitude of the primordial power spectrum at some characteristic pivot scale is given by $\Delta^2_{\GW,\text{prim}}$ is \cite{abbott1984constraints, sahni1990energy}:
\begin{equation}
	\label{eq:primordial_spectra}
	\Delta^2_{\GW, \text{prim}} \equiv {4 }\frac{k^3}{2\pi^2} |h_{\text{prim}}(k)|^2 = \frac{2}{\pi^2} \left( \frac{H_{\text{inf}}}{\mpl} \right)^2,
\end{equation}
where $\mpl = (8\pi G_N)^{-1/2}$ is the reduced Planck mass, and we adopt the same conventions of \cite{watanabe2006improved}.
Given that the factor $\sin(u+\delta)$ implicit in Eq.~\eqref{eq:spectrum_analytical} is a rapidly oscillating function, we plot and analyze only the amplitude $A(u)$ for various scenarios in what follows. The fiducial values of various cosmological parameters utilized in our calculations are provided in Table~\ref{table:param}.

\begin{table}[t]
	\centering
	\begin{tabular}{|c|c|}
    
		\hline
		Parameter & Value\\
		\hline
		$h$ & $0.7$ \\ \hline
		$\Omega_r h^2$ & $4.15\times 10^{-5}$ \\ \hline
		$\Omega_m$ & 1  \\ \hline
		$H_{\text{inf}}$ & $2.41 \times 10^{13}$ GeV \\ \hline
		$T_\text{CMB}$ & $2.34 \times 10^{-4}$ eV \\ \hline 
		$T_{\nu, \text{dec}}$ & $2$ MeV\\ \hline
	\end{tabular}
    
	\caption{Fiducial cosmological parameters adopted. Here, $h$ denotes the dimensionless Hubble parameter, $\Omega_r$ and $\Omega_m$ are relative energy densities of radiation and matter, $H_{\text{inf}}$ is the Hubble parameter at the end of inflation, $T_\text{CMB}$ is the current temperature of the CMB, and $T_{\nu, \text{dec}}$ is the neutrino decoupling temperature.
    }
    
	\label{table:param}
\end{table}

\section{Early matter domination}
\label{section:EMD}
There are multiple distinct scenarios in which one or more intervals of matter domination intersperse an initial and terminal stage of radiation domination, with the initial stage corresponding to reheating, and the terminal stage corresponding to the epoch preceding BBN~\cite{Allahverdi:2020bys}. Each scenario will result in a distinct late time spectral density depending on the number and relative duration of the matter domination phases. Nevertheless, one can infer certain general features by examining a representative example, the most notable of which is the fact that the high frequency plateau of the spectral density will be suppressed relative to the standard thermal history.      

\begin{figure}[h!]
	\centering
	\includegraphics[width=0.7\linewidth]{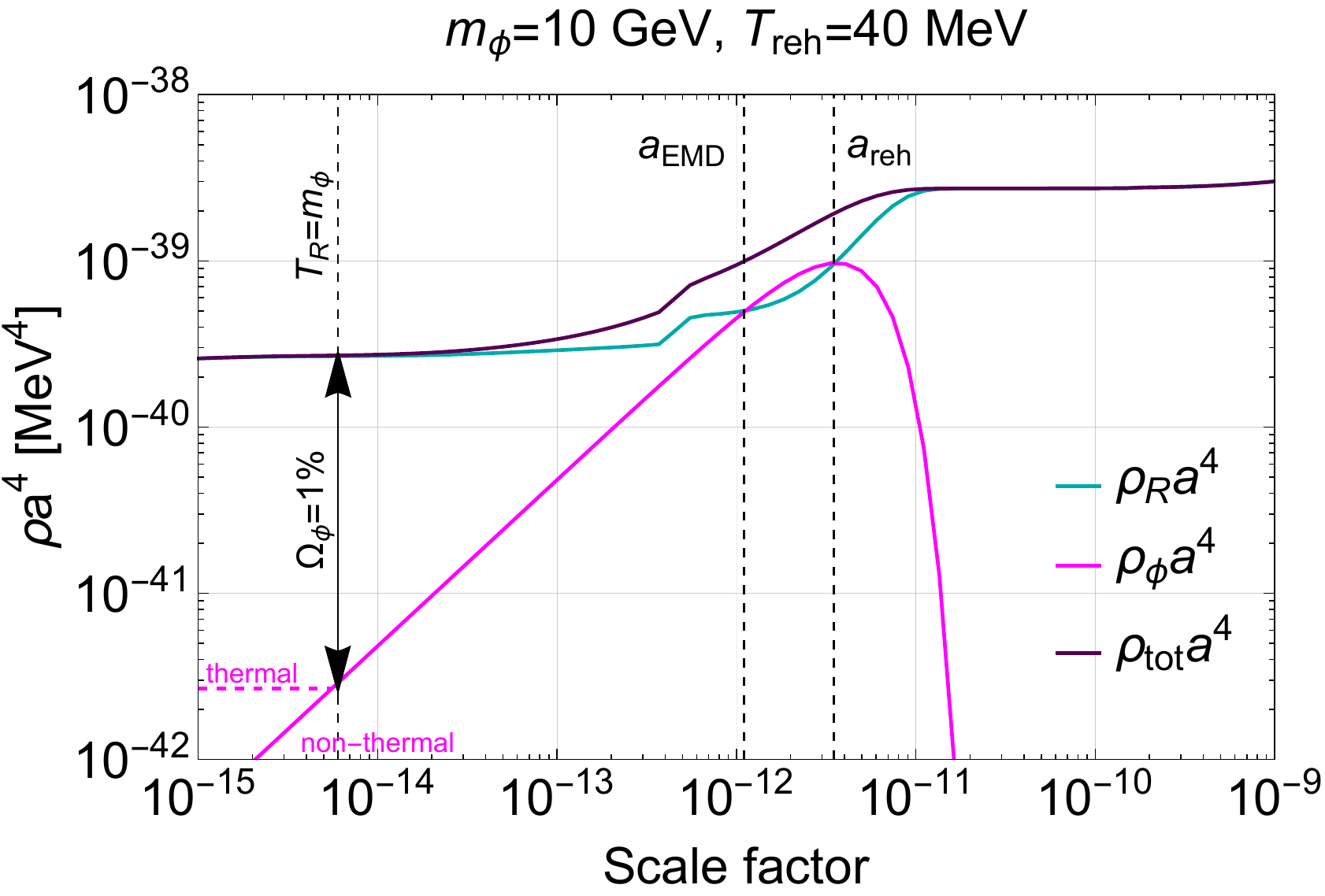}
	\caption{Evolution of the rescaled energy density $\rho a^4$ of radiation (dark cyan), $\phi$ (magenta), and total (purple) as a function of the scale factor in a model with EMD induced by $m_\phi = 10$ GeV particle with decay width $\Gamma_\phi \approx \unit[10^{-21}]{GeV}$, corresponding to reheating temperature $T_\text{reh}=40$ MeV. The dashed magenta line corresponds to a thermal $\phi$ that becomes nonrelativistic only when the temperature drops below $m_\phi$. }
	\label{fig:rho_a4_evol}
\end{figure}

One such scenario is a massive field $\phi$ that drives reheating (whether it be the inflaton itself or one of its decay products) and decays into Standard Model quanta afterwards \cite{Bernal:2019lpc, Grin:2007yg} with an intervening epoch of matter domination being the result. We assume that the universe is dominated by radiation, and $\phi$ behaves as non-relativistic matter when the plasma temperature drops below the mass of the field $T = m_\phi$, at which point the relative energy contribution of the new field is $\Omega_\phi$. In this setup, we can treat uniformly the cases of thermal and non-thermal production of $\phi$. After that, the dynamics of two dominant components, radiation and the inflaton field, is described by a system of coupled Boltzmann equations: 
\begin{align}
\label{eq:non_standard_energy_evol}
    a\frac{d\rho_{\phi}}{da} + 3 \rho_{\phi}&=-\frac{\Gamma_{\phi}\rho_{\phi}}{H},\\
\label{eq:non_standard_temp_evol}
     a\frac{ds_R}{da} + 3 s_R&=\frac{\Gamma_{\phi}\rho_{\phi}}{H T },
\end{align}
where $\rho_{\phi}$ is the energy density of $\phi$, $s_R = \frac{4\pi^2}{90} g_{\star,s}(T)T^3$ is the entropy density of radiation, related to temperature and the radiation fluid, $H$ is the Hubble parameter, and $\Gamma_\phi$ is the decay width of $\phi$ quanta, which can be expressed in terms of reheating temperature $T_{\text{reh}}$ of the plasma after the decay of $\phi$~\cite{Giudice:2000dp, Giudice:2000ex, Chung:1998rq}:
\begin{equation}
\label{eq: reheating_temp}
    \Gamma_{\phi} = \sqrt{\frac{\pi^2}{90 }g_{\star}(T_{\text{reh}})}\frac{T_{\text{reh}}^2}{\mpl},
\end{equation}
Reheating temperature is constrained by BBN: $T_{\text{reh}}\gtrsim T_{\text{BBN}}\sim 4\:\text{MeV}$ \cite{Kawasaki:2000en, Hannestad:2004px, DeBernardis:2008zz, deSalas:2015glj}. The evolution of the energy densities is shown in Fig.~\ref{fig:rho_a4_evol}.

\begin{figure}[h]
	\centering
	\includegraphics[width=0.7\linewidth]{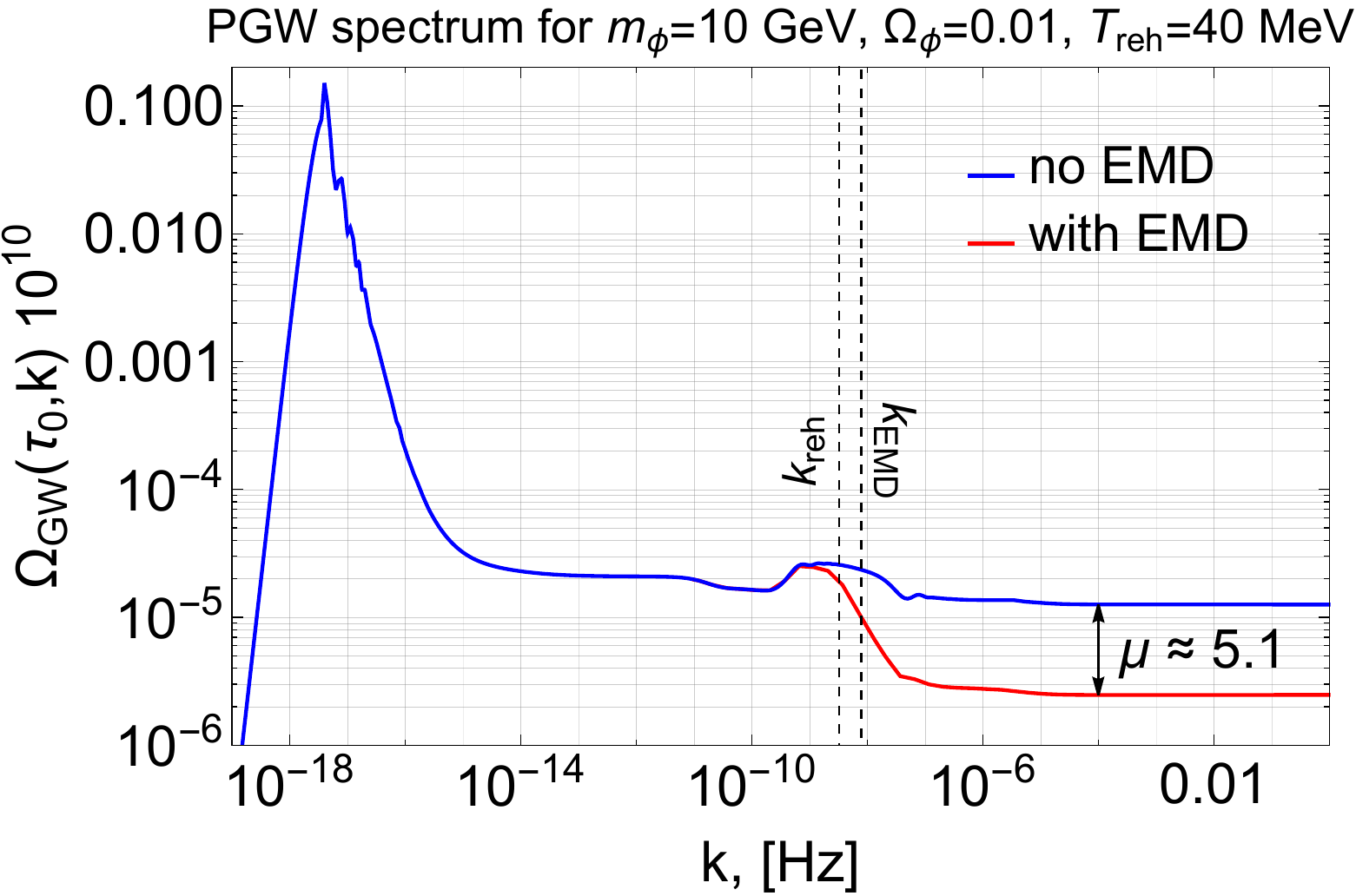}
	\caption{Spectrum of primordial gravitational waves today as a function of comoving wavelength $k$. The blue line corresponds to the conventional thermal history, the red line illustrates the EMD scenario with $m_\phi = 10$ GeV and  reheating temperature $T_\text{reh}=40$ MeV. Dashed vertical lines mark the start of the EMD (red) and reheating (blue).}
	\label{fig:SGWB_2_GeV}
\end{figure}

The period of matter domination is imprinted in the PGW spectrum in the form of a distinct step-like feature, shown in Fig.~\ref{fig:SGWB_2_GeV}. The width and height of the step can be estimated analytically. In the absence of anisotropic stress, the solution to Eq.~\eqref{eq:final_equation_text} is simply expressed in terms of Bessel functions \cite{watanabe2006improved, Saikawa:2018rcs, Kite:2021yoe}. Consequently, the energy spectrum of PGWs today can be written as:
\begin{equation}
	\label{eq:PGW_spectrum_general}
	\Omega_{\GW}(\tau_0, k) \propto [k \mathcal T_k'(k\tau_0)]^2 \propto \left[\frac{a(\tau_\text{h.c.})} {\tau_\text{h.c.}}\right]^{2} ,
\end{equation}
where $\tau_\text{h.c.} = 1/k$ is the conformal time of horizon crossing. This relation holds for a constant equation of state and for wave modes that entered the horizon far in the past $\tau_{\text{h.c.}} \ll \tau_0$. 
During RD, the scale factor behaves as $a\propto\tau$, resulting in a constant, flat spectrum. On the other hand, during MD, the relation is different $a\propto\tau^2$, resulting in a decreasing spectrum $\Omega_{\GW} \sim k^{-2}$. 
The general result for an arbitrary equation of state $\omega$ that dominates the energy density is~\cite{watanabe2006improved, Gouttenoire:2021jhk}:
\begin{equation}
	\label{eq:PGW_spectrum_special}
	\Omega_{\GW}(\tau_0, k) \sim k^{2\frac{3\omega-1}{1+3\omega}}.
\end{equation} 

Therefore, a larger decay width results in earlier reheating. Shifts at the beginning of the EMD phase can be attributed to changes in initial conditions due to rescaling to match the observed radiation energy density. The magnitude of the suppression factor $\mu$, defined as the relative height between two plateaus before and after the EMD phase, can be expressed as a function of $\phi$ parameters:
\begin{equation}
	\label{sf}
	\mu\equiv \frac{\Omega_{\text{GW}, \text{SM}}}{\Omega_{\text{GW}, \text{EMD}}}\bigg|_{k\gg k_{\text{EMD}}} \approx
    \left(\frac{\Omega_\phi m_\phi }{T_\text{reh}}\right)^{\frac{4}{3}}.
\end{equation}
We see that the longer the lifetime of the particles (i.e. the smaller the decay width) and the higher their masses, the greater the suppression of the high frequency tail. As a result, measuring the magnitude of the suppression and EMD duration could facilitate inferring the mass and decay width of particles causing EMD, potentially constraining the properties of some new physics particles (e.g., heavy neutral leptons) not accessible to current and proposed collider experiments.

\section{Long lived feebly interacting particles}

Massive feebly interacting particles (FIPs) with long lifetimes feature in several beyond the Standard Model scenarios. Examples of FIPs include dark scalars, heavy neutral portals (HNLs), dark photons, and axion-like particles (ALPs) (see~\cite{Alekhin:2015byh, Beacham:2019nyx} for an overview). Such particles have a variety of phenomenological motivations, and may even be within the experimental reach of future accelerator experiments~\cite{Mikulenko:2023olf, Mikulenko:2023iqq}. Depending on their lifetimes, they could also imprint on cosmological observables. If the FIPs in question interact feebly enough, a phase of EMD can be induced if all of the following conditions are met:
\begin{itemize}
    \item The FIPs decouple while still being present with sufficient abundance in the primordial plasma. The decoupling temperature $T_\text{fr}$ cannot be considerably lower than the FIP mass $m$, $T_\text{fr}\gtrsim m$, otherwise their number density becomes exponentially suppressed.
    \item The plasma temperature $T_\text{dec}$ at the moment of FIP decay must have dropped to be substantially below the FIP mass $T_\text{dec} \ll m$.
    \item FIP decay and subsequent thermalization temperature $T_\text{reh}$ must be above several MeV so as not to spoil the predictions of Big Bang Nucleosynthesis~\cite{Depta:2020wmr, Boyarsky:2020dzc}\footnote{Although not relevant to the scenarios we consider, FIPs that live long enough to decay in the window of redshift $z \sim 10^6$ and $10^3$ do so in an epoch when Compton scattering in the primordial plasma is not efficient enough to thermalize the decay products, imprinting a distortion of the cosmic microwave background spectrum \cite{Chluba:2013pya}.}.
\end{itemize} 

For simple order-of-magnitude estimates as to whether the described scenario occurs, we assume that the FIP production rate scales with temperature as $\Gamma_\text{prod} \sim T^{n+1}/\Lambda^{n}$ once $T\gtrsim m$. The parameters $\Lambda$, $n$ are determined by the particular phenomenological model. The time of freeze-out is given by $\Gamma(T) = H(T)$, or
\begin{equation*}
    T_\text{fr} \sim (\Lambda^{n} \mpl^{-1})^{1/(n-1)}.
\end{equation*}
On the other hand, we can make a similar estimate for the FIP decay rate by replacing the temperature with the particle mass $m$: $\Gamma_{\text{decay}} \sim m^{n+1}/\Lambda^n$. This is a reasonable assumption if there are no dominant decays into some exotic dark sector that does not contribute to the FIP production rate. Once the FIPs become nonrelativistic, the universe enters an epoch of matter-domination $H(T) \sim \sqrt{mT^3}/\mpl$, and the decay temperature reads
\begin{equation*}
    T_\text{dec} \sim m \times \left(\frac{m}{T_\text{fr}}\right)^{\frac{2}{3}(n-1)}.
\end{equation*}
The requirements for EMD -- $T_\text{dec}\ll m \lesssim T_\text{fr}$ -- can only be satisfied with $n>1$, which excludes renormalizable interactions like those corresponding to dark scalars and dark photons. However, even if a phase of EMD cannot be realized through the latter channels, changes in the number of relativistic species and bursts of anisotropic stress production can still induce imprints on the late time gravitational wave spectral density in certain scenarios. On the other hand, FIP species such as axion-like particles and heavy neutral leptons and indeed drive a phase of EMD below the electroweak scale with non-renormalizable effective interactions of the Fermi theory variety. We consider each separately.
\\

\noindent  \textbf{Axion-like particles:} We consider a model of an axion-like particle with a photon-dominant coupling (cf. the \textit{BC9} model of~\cite{Beacham:2019nyx}). Before electroweak symmetry breaking, the Lagrangian reads:
\begin{equation*}
        \mathcal{L}_{a} = \frac{1}{2}\partial_\mu a \partial^\mu a - \frac{m_a^2}{2}a^2 - \frac{g_{a\gamma\gamma}}{4\cos^2 \theta_W} aB_{\mu\nu}\tilde{B}^{\mu\nu},
\end{equation*}
with $a$ being the ALP field, and $B_{\mu\nu}$ being the field strength of the hypercharge group $\tilde B_{\mu\nu} = \frac{1}{4}\epsilon_{\mu\nu\lambda \rho} B^{\lambda \rho}$.

\begin{figure}[t!]
    \centering
    \includegraphics[width=0.8\linewidth]{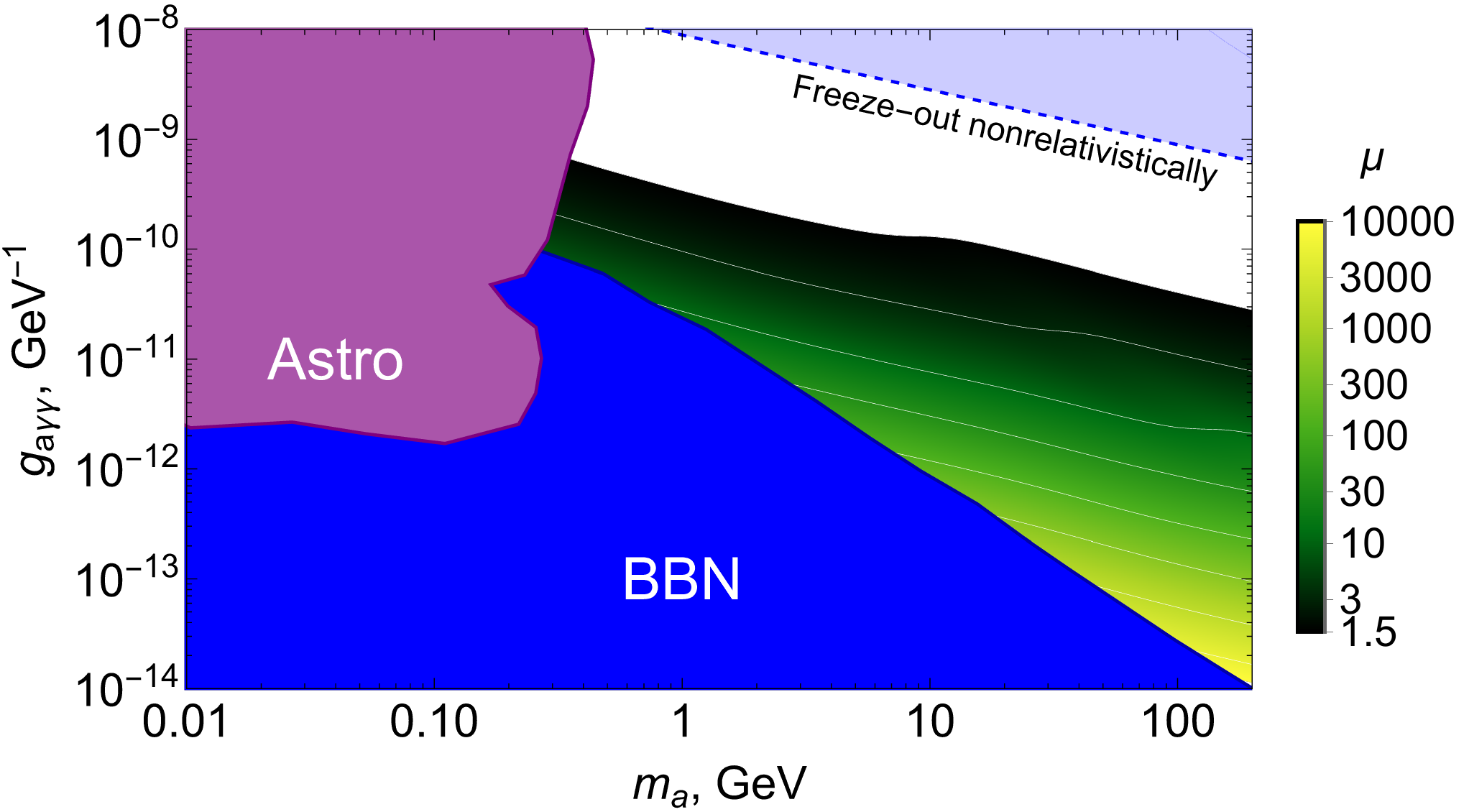}
    \caption{The suppression factor $\mu$ in the parameter space of the photon-coupled ALP with the current cosmological and astrophysical constraints~\cite{AxionLimits}. At higher $g_{a\gamma\gamma}$, interactions are too rapid, and ALPs effectively annihilate before freezing out.}
    \label{fig:axionsuppresion}
\end{figure}

As long as the ALPs decouple while they are still relativistic, the exact cross-section of processes that maintain thermal equilibrium is not relevant, and the freeze-out temperature and the axion decay width are given simply by~\cite{Depta:2020wmr}:
\begin{equation*}
    T_\text{fr} \approx 100 \left(\frac{\unit[10^{-9}]{GeV^{-1}}}{g_{a\gamma\gamma}} \right)^2 \unit{GeV},
    \qquad \Gamma(a\to \gamma \gamma) = \frac{g_{a\gamma\gamma}^2 m^3}{64\pi}. 
\end{equation*}
We numerically solve the kinetic equations to find the energy density and $a(\tau)$, assuming a thermal number density of relativistic axions. The resulting suppression scale for ALPs is then computed with Eq.~\eqref{eq:PGW_spectrum_general} and shown in Fig.~\ref{fig:axionsuppresion}.\\

\noindent \textbf{Heavy Neutral Leptons:} HNLs~\cite{Datta:2023vbs, Asaka:2020wcr,Berbig:2023yyy, Chianese:2024nyw} are an appealing extension of the Standard Model that can simultaneously account for the origin of neutrino masses via the seesaw mechanism \cite{Minkowski:1977sc, Mohapatra:1979ia, Mohapatra:1980yp, Schechter:1980gr} and the baryon asymmetry via leptogenesis \cite{Fukugita:1986hr, Asaka:2005pn, Davidson:2008bu}. The new particles mix with active neutrinos by a small mixing angle $U_\alpha$, $\alpha = e,\mu,\tau$, which couple HNLs to the standard model via a weak-like interaction. To account for the neutrino masses, the mixing angles cannot be smaller than the seesaw bound $$U^2_\text{seesaw} \gtrsim \frac{m_\nu}{m_N},$$ where $m_\nu \sim \sqrt{\Delta m^2_\text{atm}}\sim \unit[50]{meV}$ is the scale of active neutrino masses and $m_N$ is the HNL mass.

\begin{figure}[h]
	\centering
	\includegraphics[width=0.8\linewidth]{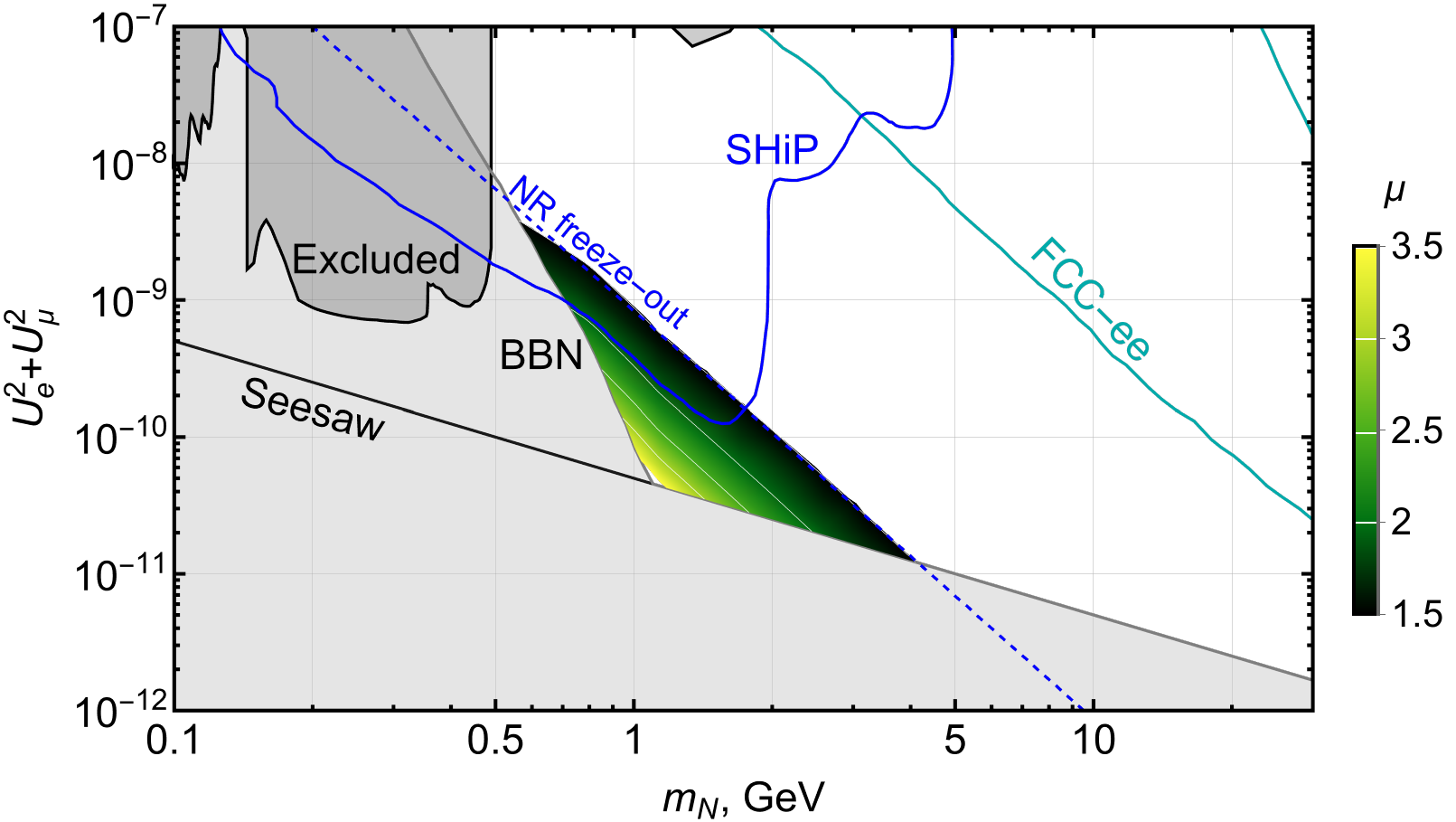}
	\caption{The suppression factor $\mu$ in the parameter space of an HNL pair with electrons mixing is presumed to be the dominant channel. Filled regions are constraints from BBN~\cite{Boyarsky:2020dzc, Bondarenko:2021cpc}, direct searches~\cite{Abdullahi:2022jlv}, and the seesaw bound, while solid lines show the projected sensitivities of SHiP~\cite{Ahdida:2023okr} and FCC-ee in the $Z$-pole mode~\cite{Blondel:2022qqo}}
	\label{fig:hnlsuppresion}
\end{figure}

Future experiments such as the planned SHiP experiment \cite{SHiP:2015vad, SHiP:2018xqw} and the proposed FCC~\cite{Antusch:2016ejd, Blondel:2022qqo,  Chrzaszcz:2020emg} aim to push the limits on the mixing angle at the GeV scale by orders of magnitude, but are incapable of closing the parameter space completely down to the seesaw bound. Therefore, complementary probes from below, such as BBN, are needed. The effect of HNLs on the spectrum of primordial GW offers a powerful probe to further constrain the unexplored parameter space.

As our findings indicate, the parameter space in which HNLs can drive an epoch of EMD covers the mass range of $\unit[1 - 10]{GeV}$, outside of which such long-lived HNLs are excluded by either the BBN constraints or the seesaw bound. In this region, HNLs are on the verge of decoupling whilst still nonrelativistic, necessitating an accurate estimation of their abundance, which we obtain via the Boltzmann equation. HNLs interact mainly via processes of the from $N + \text{SM} \leftrightarrow  \text{SM} + \text{SM}$. The rescattering channels $N+\text{SM}\leftrightarrow N+\text{SM}$ that preserve the HNL number but change its momentum require HNL-neutrino mixing to occur twice and are therefore doubly suppressed by the small mixing angle. Moreover, the weak interaction cross-section of HNLs exhibits a strong dependence on the center-of-mass energy. In light of these considerations, we opted for the following scheme: each momentum shell in the HNL phase space is treated as a separate system equilibrating with the SM thermal bath. Specifically, for the total number density $\Delta f(y)$ of HNLs in a spherical momentum shell in the radius range $(y, y+\Delta y)$, where $y = p/T$ and $p$ is the corresponding physical momentum, we solve the equation
\begin{equation*}
    \frac{d\Delta f(y)}{dt} = - 3H \Delta f - \Gamma(y T) (\Delta f - \Delta f^{\text{eq}}),
\end{equation*}
where $H$ is the Hubble parameter and $\Gamma(p) = \sum_{A \in \text{SM}} n_A \times \langle\sigma_{N+A}(p) v\rangle_\text{th. average}$ is the interaction rate summed over all species $A$ and thermally averaged. The equilibrium number density $\Delta f^{\text{eq}}$ per HNL is thus given by
\begin{equation*}
    \Delta f^\text{eq}(y) = \frac{T^3}{\pi^2} \frac{y^2 \Delta y}{\exp \sqrt{\frac{m_N^2}{T^2} + y^2} + 1}.
\end{equation*}

In our toy model, we assume that the universe is dominated by the following relativistic species at GeV temperatures: three neutrino species, electrons, muons, and light $u$, $d$, $s$. Heavier particles are highly subdominant, and the HNL contribution to the Hubble parameter is neglected. The interaction cross-section is computed precisely for each SM particle using the matrix elements from~\cite{Sabti:2020yrt} and thermally averaged. For each momentum shell $y$, we track the evolution of the spectrum until the shell freezes out. The calculated final HNL abundance serves as the initial condition for the subsequent estimate of the EMD suppression factor $\mu$ (Eq.~\eqref{sf}, plotted in Fig. \ref{fig:hnlsuppresion}).

The contours depicted on the plot are given for the model of two Majorana HNLs with equal total mixing angle $U^2$ and negligible tau neutrino mixing. In the approximation where all light particles are taken relativistic, electrons and muons are equivalent, and the results depend only on the total mixing angle $U^2_e + U^2_\mu$. We treat HNL decoupling and subsequent decay as processes, sufficiently separated in time, which is not a justified assumption since the interaction rate $\Gamma \sim G_F^2 T^5$ with the SM during decoupling competes with the decay rate $\Gamma \sim G_F^2 m_N^5$. A more accurate modeling of the full dynamics that simultaneously accounts for both effects is needed to draw definitive conclusions.\\

\noindent \textbf{Dark radiation:} FIPs that are not sufficiently long-lived to drive a period of EMD may still leave an imprint in the spectrum through two effects: the change of the effective number of DOF and the generation of anisotropic components to the stress tensor. Naively, the contribution of any such FIPs is limited to the percent level given the large value of $g_*\sim 100$ at energies above the GeV scale. In this section, we illustrate a plausible possible scenario that avoids this restriction.

\begin{figure}
	\centering
	\includegraphics[height = 0.33\linewidth]{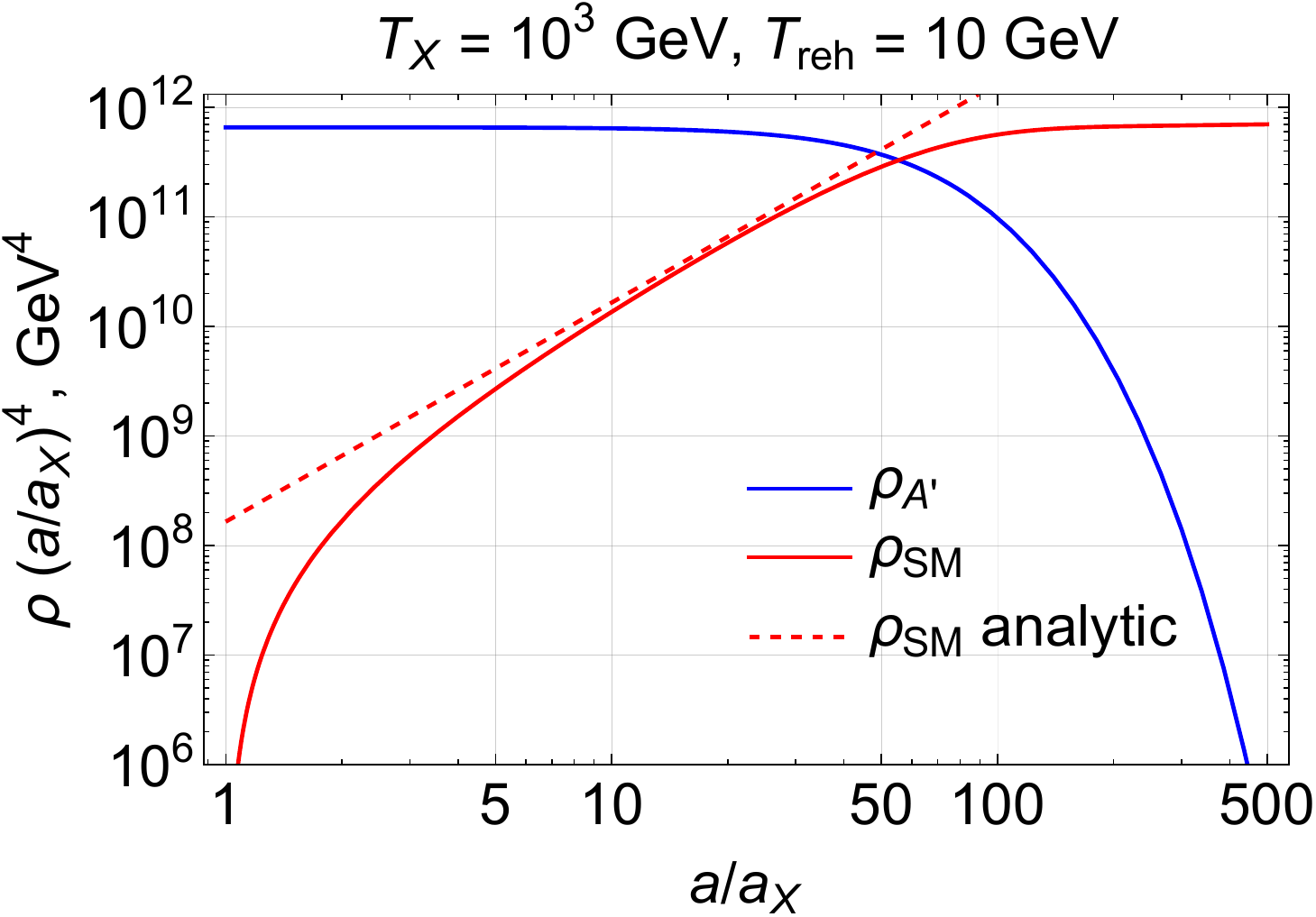}~
	\includegraphics[height = 0.33\linewidth]{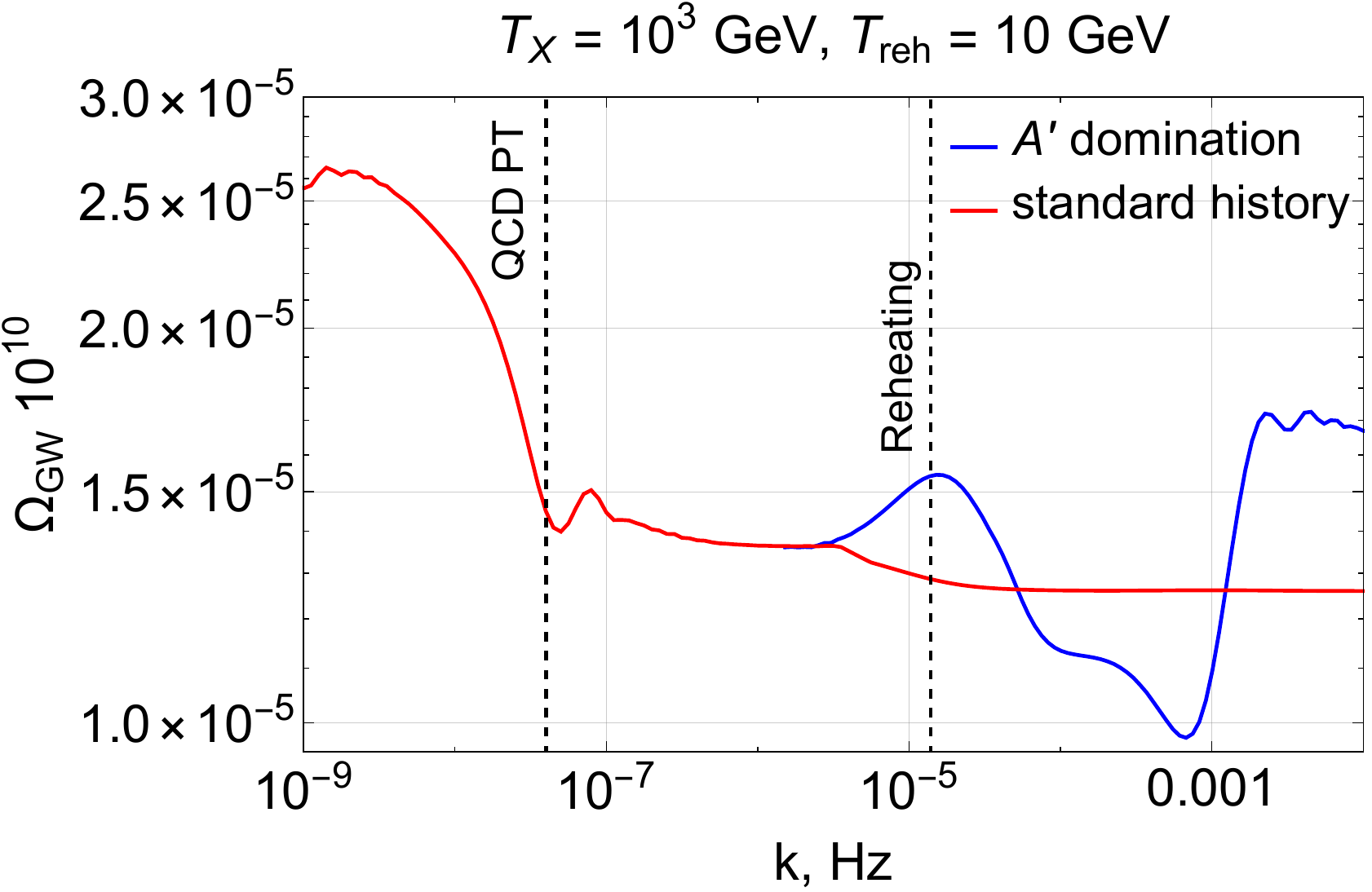}
	\caption{Left: energy transfer from dark photons into the SM. Right: the comparison of the GW spectra for the standard thermal history and early dark radiation. The height of the right plateau depends on the concrete composition of the dark sector. For the sake of concreteness, we assume that the species $X$ is a scalar and other possible particles are much heavier.}
	\label{fig:dark_photon_examples}
\end{figure}

We consider a dark sector, made up of various additional particles, that only interacts with the SM through a light mediator, e.g., dark photon or scalar. For concreteness, we assume that the mediator is the dark photon $A'$ and the Lagrangian is:
\begin{equation}
    \mathcal L = \mathcal L_\sm +\mathcal L_{\text{DS}}-\frac{1}{4} F^{'\mu\nu}F'_{\mu\nu} -  \frac{\epsilon}{2} F^{\mu\nu}F'_{\mu\nu} +\frac{m_{A'} A^{'2}}{2} + A'_\mu J^\mu_{\text{DS}}
\end{equation}
with $\mathcal L_\sm$, $\mathcal L_\text{DS}$ are Lagrangians of the Standard Model and the dark sector respectively, $F_{\mu\nu}$ and $F'_{\mu\nu}$ are the field tensors of the ordinary and dark photon, whose kinetic mixing is parameterized by the dimensionless coupling $\epsilon$. From the comparison of the equilibration rate between the dark sector and the SM $\Gamma_\text{eq} \sim \alpha \epsilon^2 T$ with the expansion rate $H\sim T^2/\mpl$, it is evident that the two sectors evolve independently at sufficiently high temperature. Therefore, it is not unreasonable to assume that the associated energy densities of the two sectors are unrelated. In our model, we make the following assumptions:
\begin{itemize}
    \item The dark sector initially dominates the energy budget of the universe
    \item The mass of the lightest `dark charged' particle $X$ of the dark sector is larger than that of dark photon $A'$
    \item Annihilation of $X$ happens before the interaction rate between $A'$ and the SM becomes sufficient to equilibrate the two sectors
\end{itemize}

After $X$ annihilates at temperature, given roughly by
\begin{equation*}
   T_{X} \sim \frac{m_X}{\ln\left[\alpha_D^2 M_\text{Pl}/m_X\right]} \sim \frac{m_X}{20},
\end{equation*}
the universe becomes populated with a gas of decoupled dark photons $A'$ and a small admixture of the SM. Rare interactions between them, such as $eA' \to e \gamma$, slowly convert the energy stored in dark photons into the SM plasma. Given that the interaction in the SM plasma is always rapid enough to keep the plasma in equilibrium, the conversion can be described via the system of equations:
\begin{align}
\label{eq:dph_eqs}
    \frac{d\rho_{A'}}{dt} =& - 4 H \rho_{A'} -  \rho_{A'} \langle n_\sm \sigma(T_{A'}, T_\sm) \rangle  \\
\label{eq:dph_eqs2}
    \frac{d\rho_\sm }{dt} =&- 3 H (\rho_\sm + p_\sm) + \rho_{A'} \cdot \langle n_\sm \sigma(T_{A'}, T_\sm)\rangle 
\end{align}
with the dark photon temperature $T_A'= T_{X} \cdot a_{X}/a$ following the standard redshift, while the SM plasma temperature $T_\sm$ is determined by its energy density $\rho_\sm \equiv \pi^2 g_* T^4/30$.
We do not include the energy transfer due to decay of dark photons, as we have found it to be a subdominant process as long as $\epsilon^2 \gtrsim m_{A'}/\mpl$. The remaining integrated collision terms are given by the averaged interaction cross-section:
\begin{equation*}
    \langle n_\sm \sigma \rangle \sim \epsilon^2 \frac{\zeta(3)}{\pi^2} T_\sm^3 \times \frac{4\pi\alpha^2}{3s} \sum_{i\in \sm} g_{n,i} Q_i^2
\end{equation*}
in which we neglect possible logarithmic enhancement factors in the cross-section and crudely assume $s \sim 4 T_\sm T_{A'}$.

The solution of Eqs.~\eqref{eq:dph_eqs}-\eqref{eq:dph_eqs2} has a number of analytic features. From any initial nonzero seed, the SM energy density quickly approaches the value
\begin{equation*}
    \rho_\sm(a) = \rho_0 \left(\frac{a_X}{a}\right)^2, \qquad \rho_0 = \left[\epsilon^2 \alpha^2 \frac{\mpl T_X}{\sqrt{6 g_*(T_\text{SM})}} \frac{\zeta(3)}{\pi}  \sum g_{n, i} Q_i^2 \right]^2
\end{equation*}
during the expansion by a factor of $a/a_X \sim e$, as shown in the left panel of Fig.~\ref{fig:dark_photon_examples}. By equating $\rho_\sm$ to $\rho_{A'} $, 
one can estimate the moment of dark photon burning and the resulting temperature of the SM plasma:
\begin{equation*}
    \frac{a_X}{a_\text{reh.}} \approx 20 \frac{\epsilon^2}{10^{-14}} \frac{\unit[1]{GeV}}{T_X}, \qquad T_\text{reh} \approx \frac{2}{g_*^{3/4}(T_\text{reh})} \frac{\epsilon^2}{10^{-14}}
\end{equation*}
To recap: in this scenario, the universe has a phase of domination by \textit{decoupled radiation} from the temperature $T_{X}$ down to the reheating temperature $T_{\text{reh}}$. Gravitational waves that enter the horizon during this epoch are affected by the anisotropic stress tensor of the decoupled $A'$ quanta, after which dark radiation gets rapidly converted into SM particles and precipitates the terminal phase of radiation domination. To quantify the damping effects of this burst of anisotropic stress production, we use the same approach as used for Eq.~\eqref{eq:final_equation_text} for free streaming neutrinos. The fraction of the energy density of decoupled radiation $f$ jumps from zero to unity at the dark radiation temperature $T_X$ and decreases afterwards. The resulting spectrum is illustrated in Figure \ref{fig:dark_photon_examples} (right panel), with two features: a growth due to the \textit{increase} of the number of degrees of freedom during reheating, and a decline due to the anisotropy effects. 

This is perhaps the most striking of all the features generated in the various scenarios we have considered in our survey -- although a number of scenarios consider the gravitational wave spectrum generated by an epoch of dark radiation \cite{Machado:2018nqk, Machado:2019xuc, Madge:2021abk, Ratzinger:2020oct, Morgante:2021bks}, bursts of anisotropic stress production can occur in various scenarios in certain regions of parameter space and have to be factored into the transfer function for accurate predictions for the late time spectral density.

\section{Kination}

\begin{figure}[h!]
	\centering
	\includegraphics[width=0.7\linewidth]{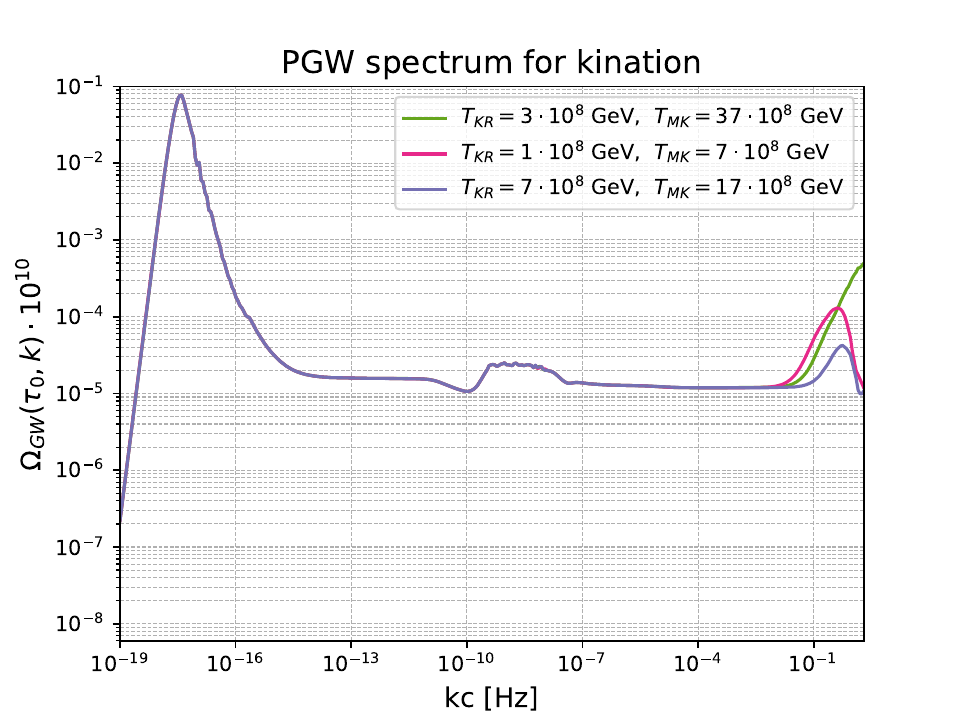}
    \caption{Spectra of primordial gravitational waves at $\tau=\tau_0$ as functions of comoving wavelength $k$ for various plausible kination scenarios. Temperatures are extracted from allowed regions shown in \cite{Co:2021lkc}.}
	\label{fig:PGW_spectrum_kination}
\end{figure}

Another plausible non-standard thermal history is provided by kination (KD), where the kinetic energy of a scalar field dominates the energy density of the universe at some epoch following on from a phase of EMD. In this section, we consider a specific model of axion kination described in \cite{Co:2019wyp, Gouttenoire:2021jhk}, where inflation is followed by Standard Model radiation domination. The contribution of the axion field $\phi$ to $\rho_{tot}$ is minor immediately after reheating, and remains constant until the expansion rate becomes smaller than the mass of the axion field $m_{S}$. At this juncture, the axion begins to oscillate around the minimum of its potential, and the spatial zero mode has the time averaged equation of state of cold dark matter, wherein a phase of EMD onsets. Kination commences once the potential energy of $\phi$ falls to below its kinetic energy. In this regime, the energy density of the Universe decreases as $\rho_{tot}\sim a^{-6}$, which is faster than that of radiation, and a terminal phase of radiation domination occurs after $\rho_{R} \approx \rho_{\phi}$. Approximating each of these regimes in a piece-wise manner \cite{Co:2021lkc}, one can show that the Hubble parameter $H(T)$ evolves as:

\begin{equation}
	\label{eq:hubble_kin}
	H(T)\sim \begin{cases}
		T^2  & \text{for}\quad \text{RD :  } T\gg T_{RM},\\
		T^2_{RM}\bigg(\frac{T}{T_{RM}}\bigg)^{ \frac{3}{2}}  & \text{for}\quad \text{MD :  } T_{RM}\gg T \gg T_{MK},\\
		T^2_{KR}\bigg(\frac{T}{T_{KR}}\bigg)^{ 3}  & \text{for}\quad\text{KD :  } T_{MK}\gg T \gg T_{KR},\\
		T^2 & \text{for}\quad\text{RD :  } T_{KR}\gg T,
	\end{cases}
\end{equation}
Where $T_{RM}$ is the temperature at the beginning of EMD, $T_{MK}$ is the temperature at the beginning of KD, and $T_{KR}$ is the temperature at the end of kination epoch.

The allowed window for kination is quite large. Because EMD ends without entropy injection, EMD and KD can occur even after the BBN, although it must still occur before recombination. The cosmological constraints on the $T_{RM}$ and $T_{KR}$ are provided in \cite{Co:2021lkc, Gouttenoire:2021jhk}.

To determine the PGW spectrum, we utilize the piece-wise approximation Eq.~\eqref{eq:hubble_kin} and keep all the standard cosmological parameters as indicated in Table~\ref{table:param}. Values of $T_{RM}$ and $T_{KR}$ were extracted from \cite{Co:2021lkc}, while $T_{MK}$ was calculated via
\begin{equation}
	\label{eq:TMK}
	T_{MK} =  T_{RM}^{\frac{1}{3}} T_{KR}^{\frac{2}{3}}. 
\end{equation}
We evaluated the spectral density only for early kination (occurring before BBN) and used temperatures allowing lepto-ALPgenesis with various axion field masses $m_{S}$ and vacuum field values $f_{a}$. Those parameters are chosen to explain the observed baryon asymmetry by lepto-ALPgenesis for illustrative purposes. The results are shown in Figure~\ref{fig:PGW_spectrum_kination} where a clear peak in the PGW spectrum is evident, consistent with the findings of \cite{Co:2021lkc, Gouttenoire:2021jhk}. The relation Eq.~\eqref{eq:TMK} ensures that the high frequency plateau that onsets after the peak in the spectrum will be at the same power as the plateau for modes that re-enter the horizon during the terminal stage of radiation domination. It has been noted that in principle, an enhanced high frequency plateau can result if the equation of state parameter transitions from the initial phase of radiation domination post-reheating directly to a phase of kination \cite{Soman:2024zor}. It would be interesting to investigate the spectrum that would result from a background model construction that consistently actualizes this.

\section{Concluding Remarks}

This investigation has surveyed the implications for the late time spectral density of gravitational waves for a range of phenomenologically motivated scenarios of non-standard thermal histories. In addition to corroborating the findings of a range of studies via a detailed transfer function analysis, we have also uncovered the novel possibility of damping at intermediate scales via bursts of anisotropic stress production in scenarios involving dark photon production. Realistic detection prospects, either through pulsar timing arrays \cite{10.1093/nsr/nwx126, Lommen:2015gbz, Reardon:2023gzh} or next generation interferometry \cite{Cahillane:2022pqm, ET:2019dnz} hinge on an enhanced signal at the relevant frequency ranges, and hence only meaningfully place scenarios involving kination, first order phase transitions, or secondary gravitational wave production within reach of near term observations. Nevertheless, it is informative to place targets on longer term \cite{Crowder:2005nr}, possibly even tabletop experimental efforts \cite{Aggarwal:2020olq}. That it is not just in principle, but rather practically possible to directly probe cosmological epochs that precede nucleosynthesis is a remarkable feature of gravitational wave cosmology, one that we hope will be in the realm of the actual within the lifetime of some of our readers.

\section{Acknowledgments}
We wish to thank Jens Chluba, Wolfram Ratzinger, and Pedro Schwaller for their comments on the manuscript. OM acknowledges funding from the European Research Council (ERC) under the European Union's Horizon 2020 research and innovation programme (GA 694896), from the NWO Physics Vrij Programme ``The Hidden Universe of Weakly Interacting Particles'', No. 680.92.18.03, which is partly financed by the Dutch Research Council NWO.

\bibliographystyle{unsrt}
\bibliography{Bib}

\end{document}